
\magnification \magstep1
\raggedbottom
\openup 4\jot
\voffset6truemm
\rightline {DSF preprint 95/29, June 1995}
\centerline {\bf ASYMPTOTIC HEAT KERNELS IN}
\centerline {\bf QUANTUM FIELD THEORY}
\vskip 1cm
\centerline {\bf Giampiero Esposito$^{1,2}$}
\vskip 1cm
\noindent
{\it ${ }^{1}$Istituto Nazionale di Fisica Nucleare,
Sezione di Napoli, Mostra d'Oltremare Padiglione 20,
80125 Napoli, Italy;}
\vskip 0.3cm
\noindent
{\it ${ }^{2}$Dipartimento di Scienze Fisiche,
Mostra d'Oltremare Padiglione 19, 80125 Napoli, Italy.}
\vskip 1cm
\noindent
{\bf Abstract.} Asymptotic expansions were first introduced by
Henri Poincar\'e in 1886. This paper describes their application
to the semi-classical evaluation of amplitudes in quantum field
theory with boundaries.
By using zeta-function regularization, the conformal
anomaly for a massless spin-${1\over 2}$ field in flat Euclidean
backgrounds with boundary is obtained on imposing locally
supersymmetric boundary conditions. The quantization program for
gauge fields and gravitation in the presence of boundaries is then
introduced by focusing on conformal anomalies for higher-spin
fields. The conditions under which the covariant Schwinger-DeWitt
and the non-covariant, mode-by-mode analysis of quantum amplitudes
agree are described.
\vskip 0.8cm
\noindent
To appear in: Proceedings of the Henri Poincar\'e
Conference, Protvino, June 1994.
\vskip 100cm
\leftline {\bf 1. Introduction}
\vskip 1cm
\noindent
In 1886, Henri Poincar\'e published a paper on the irregular
integrals of linear equations [1]. Section I of [1] is devoted
to the asymptotic series, and Poincar\'e begins by discussing the
peculiar properties of Stirling's series:
$$
\log \Gamma(x+1)={1\over 2} \log(2 \pi)
+ \Bigr(x+{1\over 2}\Bigr)\log(x) -x
+{B_{1}\over 1 \cdot 2}{1\over x}
-{B_{2}\over 3 \cdot 4}{1\over x^{2}}
+{B_{3}\over 5 \cdot 6}{1\over x^{3}}
- ...
\; \; \; \; .
\eqno (1.1)
$$
Poincar\'e points out that this series is always diverging,
but one can use it at large $x$. What happens is that, after
decreasing very rapidly, the terms become unboundedly large.
Nevertheless, if one takes the smallest term, the corresponding
error in the evaluation of $\log \Gamma(x+1)$ is very small.
These properties lead to the following definitions, hereafter
presented in the more general case of functions defined on a
subset of complex numbers.

Let $f$ be a function defined in an unbounded domain $\Omega$.
A power series $\sum_{n=0}^{\infty}a_{n}z^{-n}$,
{\it converging} or {\it diverging}, is said to be an asymptotic
expansion of $f$ if, $\forall$ fixed $N \geq 0$, one has
$$
f(z)=\sum_{n=0}^{N}a_{n}z^{-n}+{\rm O}\Bigr(z^{-(N+1)}\Bigr)
\; \; \; \; ,
\eqno (1.2)
$$
as $z \rightarrow \infty$. Hence one finds
$$
\lim_{z \to \infty} z^{N} \mid f(z)-S_{N}(z)\mid =0
\; \; \; \; ,
\eqno (1.3)
$$
where $S_{N}$ is the sum of the first $N+1$ terms of the series,
and one writes
$$
f(z) \sim \sum_{n=0}^{\infty} a_{n}z^{-n}
\; \; \; \; {\rm as} \; \; \; \; z \rightarrow \infty
\; \; \; \; .
\eqno (1.4)
$$
The asymptotic expansions (hereafter denoted by A.E.) have some
basic properties, which are standard (by now) but very useful.
They are as follows.
\vskip 0.3cm
\noindent
(i) The A.E. of $f$, if it exists, is unique.
\vskip 0.3cm
\noindent
(ii) A.E. may be summed, i.e. if $f(z) \sim \sum_{n=0}^{\infty}
a_{n}z^{-n}$ and $g(z) \sim \sum_{n=0}^{\infty} b_{n}z^{-n}$,
then
$$
\alpha f(z) + \beta g(z) \sim \sum_{n=0}^{\infty}
\Bigr(\alpha \; a_{n}+\beta \; b_{n}\Bigr)z^{-n}
\; \; \; \; ,
\eqno (1.5)
$$
as $z \rightarrow \infty$ in $\Omega$.
\vskip 0.3cm
\noindent
(iii) A.E. can be multiplied, i.e.
$$
f(z)g(z) \sim \sum_{n=0}^{\infty}c_{n}z^{-n}
\; \; \; \; ,
\eqno (1.6)
$$
where $c_{n} \equiv \sum_{s=0}^{n}a_{s} \; b_{n-s}$.
\vskip 0.3cm
\noindent
(iv) If $f$ is continuous in the domain $\Omega$ defined by
$\mid z \mid > a$, ${\rm arg}(z) \in [\theta_{0},\theta_{1}]$,
and if (1.4) holds, then
$$
\int_{z}^{\infty}\biggr[f(t)-a_{0}-{a_{1}\over t}\biggr]dt
\sim \sum_{n=1}^{\infty} {a_{n+1}\over n} \; z^{-n}
\; \; \; \; ,
\eqno (1.7)
$$
as $z \rightarrow \infty$ in $\Omega$, where the integration is
taken along a line $z \rightarrow \infty$ with fixed argument.
This is what one means by term-by-term integration of A.E.
\vskip 0.3cm
\noindent
(v) Term-by-term differentiation can also be performed, providing
in the domain $\Omega$ defined by $\mid z \mid > R$,
${\rm arg}(z) \in ]\theta_{0},\theta_{1}[$, the function $f$
satisfying (1.4) has continuous derivative $f'$, and $f'$ has an
A.E. as $z \rightarrow \infty$ in $\Omega$. Then
$$
f'(z) \sim -\sum_{n=1}^{\infty}n \; a_{n} \; z^{-(n+1)}
\; \; \; \; {\rm as} \; \; \; \; z \rightarrow \infty
\; \; \; \; {\rm in} \; \; \; \; \Omega
\; \; \; \; .
\eqno (1.8)
$$
\vskip 1cm
\leftline {\bf 2. Zeta-function and heat kernels}
\vskip 1cm
\noindent
We are here interested in the approach to quantum field theory
in terms of Feynman path integrals. Hence we study the amplitudes
of going from data on a spacelike surface $\Sigma_{1}$ to data
on a spacelike surface $\Sigma_{2}$. For example,
in the case of real scalar fields $\phi$ in a curved background
$M$, the data are the induced 3-metric $h$ and a linear
combination of $\phi$ and its normal derivative:
$a \phi + b {\partial \phi \over \partial n}$. The latter
reduces to homogeneous Dirichlet conditions if $b=0$, and
Neumann conditions if $a=0$. Otherwise, it is a Robin boundary
condition. The quantum amplitudes are functionals of these
boundary data. On making a Wick rotation and
using the background-field method, one
expands both the 4-metric $g$ and the field $\phi$ around
solutions of the classical field equations as
$g=g_{0}+{\overline g}$ and $\phi=\phi_{0}+{\overline \phi}$.
If second-order cross-terms vanish in the Euclidean action
$I_{E}$, the logarithm of the quantum amplitude $Z$ takes the
asymptotic form [2]
$$
\log(Z) \sim -I_{E}(g_{0})+\log \int \mu_{1}[{\overline \phi}]
e^{-I_{2}[{\overline \phi}]}
+ \log \int \mu_{2}[{\overline g}]e^{-I_{2}[{\overline g}]}
\; \; \; \; ,
\eqno (2.1)
$$
where $\mu_{1}$ and $\mu_{2}$ are suitable measures on the spaces
of scalar-field and metric perturbations, respectively. The part
$I_{2}[{\overline \phi}]$ of the action which is quadratic in
scalar-field perturbations involves a second-order elliptic
operator $\cal B$. Assuming completeness of the set
$\{ \varphi_{n} \}$ of eigenfunctions of $\cal B$, with
eigenvalues $\lambda_{n}$, the corresponding contribution to
one-loop quantum amplitudes involves an infinite product of
Gaussian integrals, i.e.
$$
\prod_{n=n_{0}}^{\infty}
\int \mu \; dy_{n} \; e^{-{\lambda_{n}\over 2} \; y_{n}^{2}}
={1\over \sqrt{{\rm det} \biggr({1\over 2}\pi^{-1}\mu^{-2}
{\cal B} \biggr)}}
\; \; \; \; .
$$
To make sense of this infinite product of eigenvalues, one is
thus led to use zeta-function regularization. This is a
rigorous mathematical tool which relies on the spectral theorem,
according to which for any elliptic, self-adjoint, positive-definite
operator $\cal A$, its complex powers ${\cal A}^{-s}$ can be
defined. Hence its zeta-function is defined as
$$
\zeta_{\cal A}(s) \equiv {\rm Tr}\Bigr[{\cal A}^{-s}\Bigr]
=\sum_{\lambda >0} \lambda^{-s}
\; \; \; \; ,
\eqno (2.2)
$$
where the eigenvalues in (2.2) are counted with their degeneracies.
If $n$ is the dimension of our Riemannian manifold $M$, and $m$
is the order of $\cal A$, its zeta-function has an analytic
continuation to the whole complex-$s$ plane as a meromorphic
function, given by [3-4]
$$
``\zeta_{\cal A}(s)"=\sum_{k \not = 0, k=-n}^{N}
{a_{k}\over \Bigr(s+{k\over m}\Bigr)}+H_{N}
\; \; \; \; ,
\eqno (2.3)
$$
where $H_{N}$ is holomorphic for $Re(s)>-{N\over m}$.
Thus, on using analytic continuations, $\zeta_{\cal A}(0)$ is
actually finite, and its value gives information about one-loop
divergences of physical theories and scaling properties of
quantum amplitudes. The relation
${\rm det}{\cal A}=e^{-\zeta'(0)}$, first obtained by formal
differentiation, is then used to define ${\rm det}{\cal A}$
after performing the suitable analytic continuation.

Coming back to the operator $\cal B$ in the Euclidean action
for real scalar fields, one now begins by studying the
corresponding heat equation, whose Green's function reads
$$
F(x,y,t)=\sum_{n=n_{0}}^{\infty} \sum_{m=m_{0}}^{\infty}
e^{-\lambda_{n,m} t}
\; \varphi_{n,m}(x) \otimes \varphi_{n,m}(y)
\; \; \; \; .
\eqno (2.4)
$$
The corresponding integrated heat kernel, defined by
$$
G(t) \equiv \int_{M}{\rm Tr} \; F \; \sqrt{{\rm det} \; g_{0}}
\; d^{4}x
\; \; \; \; ,
\eqno (2.5)
$$
is then related to the zeta-function by an inverse Mellin transform:
$$
\zeta(s)={1\over \Gamma(s)}\int_{0}^{\infty}
t^{s-1}G(t) \; dt
\; \; \; \; .
\eqno (2.6)
$$
Since, as $t \rightarrow 0^{+}$, $G(t)$ has an A.E. in the form
$$
G(t) \sim b_{0}t^{-2}+b_{1}t^{-{3\over 2}}+b_{2}t^{-1}
+b_{3}t^{-{1\over 2}}+b_{4}+{\rm O}(\sqrt{t})
\; \; \; \; ,
\eqno (2.7)
$$
whenever the boundary conditions ensure self-adjointness of
$\cal B$, the $\zeta(0)$ value is equal to the constant
coefficient $b_{4}$ appearing in (2.7).

To understand the applications presented in section 3, it is
also necessary to define the zeta-function at large $x$, i.e.
$$
\zeta \Bigr(s,x^{2}\Bigr) \equiv
\sum_{n=n_{0}}^{\infty} \sum_{m=m_{0}}^{\infty}
{\Bigr(\lambda_{n,m}+x^{2}\Bigr)}^{-s}
\; \; \; \; .
\eqno (2.8)
$$
The A.E. (2.7) is here re-written as
$$
G(t) \sim \sum_{n=0}^{\infty}B_{n} t^{{n\over 2}-2}
\; \; \; \; t \rightarrow 0^{+}
\; \; \; \; .
\eqno (2.9)
$$
For problems with boundaries, the eigenfunctions are usually
expressed in terms of Bessel functions. By virtue of the
boundary conditions, a linear (or non-linear) combination of
Bessel functions is set to zero. Denoting by $F_{p}$ the
function occurring in this eigenvalue condition, one has the
identity [2,5]
$$
\Gamma(3) \zeta(3,x^{2})=\sum_{p=0}^{\infty}
N_{p}{\biggr({1\over 2x}{d\over dx}\biggr)}^{3}
\log \Bigr[(ix)^{-p} F_{p}(ix)\Bigr]
\; \; \; \; ,
\eqno (2.10)
$$
where $N_{p}$ is the corresponding degeneracy. On the other
hand, by virtue of (2.9) one finds
$$
\Gamma(3) \zeta(3,x^{2})=\int_{0}^{\infty}t^{2}
e^{-x^{2}t} \; G(t) \; dt \sim
\sum_{n=0}^{\infty}B_{n} \Gamma \Bigr(1+{n\over 2}\Bigr)
x^{-n-2} \; \; \; \; .
\eqno (2.11)
$$
Thus, by comparison, one finds that $\zeta(0)=B_{4}$ is half
the coefficient of $x^{-6}$ in the uniform A.E. of the
right-hand side of (2.10).
\vskip 1cm
\leftline {\bf 3. Conformal anomalies for massless
spin-${1\over 2}$ fields}
\vskip 1cm
\noindent
The analysis of boundary conditions in quantum field theory
has motivated the introduction of locally supersymmetric
boundary conditions for bosonic and fermionic fields. We here
focus on a massless fermionic field at one-loop about a flat
Euclidean background bounded by a 3-sphere, following [2,6].
Using 2-component spinor notation, such a field is expressed
by a pair of independent spinor fields $\psi^{A}$ and
${\widetilde \psi}^{A'}$. Their expansion on a family of
3-spheres centred on the origin can be written as
($\tau$ being the Euclidean-time coordinate)
$$
\psi^{A}=
{\tau^{-{3\over 2}}\over 2\pi}
\sum_{n=0}^{\infty}\sum_{p,q=1}^{(n+1)(n+2)}
\alpha_{n}^{pq}
\Bigr[m_{np}(\tau)\rho^{nqA}
+{\widetilde r}_{np}(\tau){\overline \sigma}^{nqA}\Bigr]
\; \; \; \; ,
\eqno (3.1)
$$
$$
{\widetilde \psi}^{A'}=
{\tau^{-{3\over 2}}\over 2\pi}
\sum_{n=0}^{\infty}\sum_{p,q=1}^{(n+1)(n+2)}
\alpha_{n}^{pq}
\Bigr[{\widetilde m}_{np}(\tau){\overline \rho}^{nqA'}
+r_{np}(\tau)\sigma^{nqA'}\Bigr]
\; \; \; \; .
\eqno (3.2)
$$
With our notation, the $\alpha_{n}^{pq}$ are block-diagonal
matrices with blocks $\pmatrix {1&1 \cr 1&-1 \cr}$, and the
$\rho$- and $\sigma$-harmonics obey the identities described
in [2,6]. Our boundary conditions are
$$
\sqrt{2} \; {_{e}n_{A}^{\; \; A'}} \;
\psi^{A}=\epsilon \; {\widetilde \psi}^{A'}
\; \; \; \; {\rm on} \; \; \; \; S^{3}
\; \; \; \; ,
\eqno (3.3)
$$
where $\epsilon \equiv \pm 1$, and ${_{e}n_{A}^{\; \; A'}}$ is
the Euclidean normal to $S^{3}$ [2,6]. As shown in [2,6], the
corresponding eigenvalue condition is found to be
$$
F(E) \equiv {\Bigr[J_{n+1}(E)\Bigr]}^{2}
-{\Bigr[J_{n+2}(E)\Bigr]}^{2}=0
\; \; \; \; \forall n \geq 0
\; \; \; \; .
\eqno (3.4)
$$
Remarkably, the function $F$ occurring in (3.4) admits a
canonical-product representation in terms of its eigenvalues
$\mu_{i}$ as ($\gamma$ being a constant)
$$
F(z)=\gamma \; z^{2(n+1)} \prod_{i=1}^{\infty}
\biggr(1-{z^{2}\over \mu_{i}^{2}}\biggr)
\; \; \; \; .
\eqno (3.5)
$$
Thus, setting $m \equiv n+2$, one finds [2,6]
$$
J_{m-1}^{2}(x)-J_{m}^{2}(x)={J_{m}'}^{2}
+\biggr({m^{2}\over x^{2}}-1\biggr)J_{m}^{2}
+2{m\over x}J_{m} J_{m}'
\; \; \; \; .
\eqno (3.6)
$$
Thus, on making the analytic continuation
$x \rightarrow ix$ and then defining
$\alpha_{m} \equiv \sqrt{m^{2}+x^{2}}$, one obtains [2,6]
$$ \eqalignno{
\log \biggr[(ix)^{-2(m-1)}\Bigr(J_{m-1}^{2}-J_{m}^{2}\Bigr)
(ix)\biggr]
& \sim -\log(2\pi) + \log(\alpha_{m})+2\alpha_{m} \cr
&-2m \log(m+\alpha_{m})+\log({\widetilde \Sigma})
\; \; \; \; .
&(3.7)\cr}
$$
In the A.E. (3.7), $\log({\widetilde \Sigma})$ admits
an asymptotic series in the form
$$
\log ({\widetilde \Sigma}) \sim
\biggr[\log(c_{0})+{A_{1}\over \alpha_{m}}
+{A_{2}\over \alpha_{m}^{2}}
+{A_{3}\over \alpha_{m}^{3}}
+... \biggr]
\; \; \; \; ,
\eqno (3.8)
$$
where, on using the Debye polynomials for uniform A.E. of Bessel
functions [2], one finds (hereafter $t \equiv {m\over \alpha_{m}}$)
$$
c_{0}=2(1+t)
\; \; \; \; ,
\eqno (3.9)
$$
$$
A_{1}=\sum_{r=0}^{2}k_{1r} t^{r}
\; \; \; \; , \; \; \; \;
A_{2}=\sum_{r=0}^{4}k_{2r}t^{r}
\; \; \; \; , \; \; \; \;
A_{3}=\sum_{r=0}^{6}k_{3r}t^{r}
\; \; \; \; ,
\eqno (3.10)
$$
where [2,6]
$$
k_{10}=-{1\over 4}
\; \; \; \; k_{11}=0 \; \; \; \; k_{12}={1\over 12}
\; \; \; \; ,
\eqno (3.11)
$$
$$
k_{20}=0
\; \; \; \; k_{21}=-{1\over 8} \; \; \; \;
k_{22}=k_{23}={1\over 8} \; \; \; \;
k_{24}=-{1\over 8}
\; \; \; \; ,
\eqno (3.12)
$$
$$
k_{30}={5\over 192}
\; \; \; \; k_{31}=-{1\over 8} \; \; \; \;
k_{32}={9\over 320} \; \; \; \;
k_{33}={1\over 2}
\; \; \; \; ,
\eqno (3.13)
$$
$$
k_{34}=-{23\over 64}
\; \; \; \; k_{35}=-{3\over 8} \; \; \; \;
k_{36}={179\over 576}
\; \; \; \; .
\eqno (3.14)
$$
The corresponding zeta-function at large (cf. section 2)
has a uniform A.E. given by
$$
\Gamma(3)\zeta(3,x^{2}) \sim W_{\infty}
+\sum_{n=5}^{\infty}{\hat q}_{n} x^{-2-n}
\; \; \; \; ,
\eqno (3.15)
$$
where, defining
$$
S_{1}(m,\alpha_{m}(x)) \equiv -\log(\pi)+2\alpha_{m}
\; \; \; \; ,
\eqno (3.16)
$$
$$
S_{2}(m,\alpha_{m}(x)) \equiv -(2m-1)\log(m+\alpha_{m})
\; \; \; \; ,
\eqno (3.17)
$$
$$
S_{3}(m,\alpha_{m}(x)) \equiv \sum_{r=0}^{2}k_{1r} \; m^{r}
\alpha_{m}^{-r-1}
\; \; \; \; ,
\eqno (3.18)
$$
$$
S_{4}(m,\alpha_{m}(x)) \equiv \sum_{r=0}^{4}
k_{2r} \; m^{r} \alpha_{m}^{-r-2}
\; \; \; \; ,
\eqno (3.19)
$$
$$
S_{5}(m,\alpha_{m}(x)) \equiv \sum_{r=0}^{6}
k_{3r} \; m^{r} \alpha_{m}^{-r-3}
\; \; \; \; ,
\eqno (3.20)
$$
$W_{\infty}$ can be obtained as [2,6]
$$
W_{\infty}=\sum_{m=0}^{\infty}\Bigr(m^{2}-m\Bigr)
{\biggr({1\over 2x}{d\over dx}\biggr)}^{3}
\left[\sum_{i=1}^{5}S_{i}(m,\alpha_{m}(x))\right]
\; \; \; \; .
\eqno (3.21)
$$
The resulting $\zeta(0)$ value receives contributions from
$S_{2},S_{4}$ and $S_{5}$ only, and is given by [2,6]
$$
\zeta(0)=-{1\over 120}+{1\over 24}+{1\over 2}\sum_{r=0}^{4}k_{2r}
-{1\over 2}\sum_{r=0}^{6}k_{3r}={11\over 360}
\; \; \; \; .
\eqno (3.22)
$$
Of course, for a massless Dirac field, the full $\zeta(0)$ is
twice the value in (3.22):
$$
\zeta_{{\rm Dirac}}(0)={11\over 180}
\; \; \; \; .
\eqno (3.23)
$$
\vskip 1cm
\leftline {\bf 4. The BKKM function}
\vskip 1cm
\noindent
So far, the most powerful algorithm for direct $\zeta(0)$
calculations is the one described and applied in [7-9], since
it does not rely on the knowledge of the many coefficients
appearing in the Debye polynomials for uniform A.E.
of Bessel functions. With the notation in [7-9], one writes
$f_{n}(M^{2})$ for the function occurring in the equation
obeyed by the eigenvalues by virtue of boundary conditions,
and $d(n)$ for the degeneracy of the eigenvalues. One then
defines the BKKM function [7-9]
$$
I(M^{2},s) \equiv \sum_{n=n_{0}}^{\infty} d(n) \; n^{-2s}
\; \log \Bigr[f_{n}(M^{2})\Bigr]
\; \; \; \; .
\eqno (4.1)
$$
Such a function has an analytic continuation to the whole
complex-$s$ plane as a meromorphic function, i.e.
$$
``I(M^{2},s)"={I_{\rm pole}(M^{2})\over s}
+I^{R}(M^{2})+{\rm O}(s)
\; \; \; \; .
\eqno (4.2)
$$
The $\zeta(0)$ value is then obtained as
$$
\zeta(0)=I_{\rm log}+I_{\rm pole}(\infty)
-I_{\rm pole}(0)
\; \; \; \; ,
\eqno (4.3)
$$
where $I_{\rm log}=I_{\rm log}^{R}$ is the coefficient of
$\log(M)$ from $I(M^{2},s)$ as $M \rightarrow \infty$,
and $I_{\rm pole}(M^{2})$ is the residue at $s=0$.
Remarkably, $I_{\rm log}$ and $I_{\rm pole}(\infty)$ are
obtained from the uniform A.E. of modified Bessel functions
as their order tends to $\infty$ and $M \rightarrow \infty$,
while $I_{\rm pole}(0)$ is obtained from the limiting behaviour
of such Bessel functions as $M \rightarrow 0$.
\vskip 1cm
\leftline {\bf 5. Recent results}
\vskip 1cm
\noindent
The calculation outlined in section 3 is just an example of the
many difficult calculations of conformal anomalies in the
presence of boundaries appearing in the recent literature.
The same $\zeta(0)$ value has been obtained by using the even more
powerful technique elaborated in [7], as shown in [8-9]. The
motivations for this analysis come from the quantization of
closed cosmologies, from perturbative supergravity, and from
the need to get a better understanding of different quantization
techniques in field theory (i.e. reduction to physical degrees
of freedom before quantization, or Faddeev-Popov technique, or
Batalin-Fradkin-Vilkovisky method).
Here we summarize the recent results
for bosonic fields on using {\it relativistic} gauges within the
Faddeev-Popov formalism [10-12].
\vskip 0.3cm
\noindent
(5.1) In the Lorentz gauge, the mode-by-mode analysis of
one-loop amplitudes for vacuum Maxwell theory agrees with the
results of the Schwinger-DeWitt technique, both in the
one-boundary case (the disk) and in the two-boundary case
(the ring).
\vskip 0.3cm
\noindent
(5.2) In the presence of boundaries, the effects of gauge modes
and ghost modes {\it do not} cancel each other.
\vskip 0.3cm
\noindent
(5.3) When combined with the contribution of physical degrees of
freedom, this lack of cancellation is exactly what one needs
to achieve agreement with the results of the Schwinger-DeWitt
technique. Thus, physical degrees of freedom are, by themselves,
insufficient to recover the full information about one-loop
amplitudes.
\vskip 0.3cm
\noindent
(5.4) Even on taking into account physical, non-physical and
ghost modes, the analysis of relativistic gauges different
from the Lorentz gauge yields gauge-invariant amplitudes
only in the two-boundary case.
\vskip 0.3cm
\noindent
(5.5) The conditions under which one can decouple gauge
modes in the presence of boundaries have been characterized.
\vskip 0.3cm
\noindent
(5.6) Changing the gauge leads to a continuous, multi-paramater
variation of a matrix of elliptic self-adjoint operators.
Out of the eigenvalues of such operators one can obtain a
meromorphic function whose residue at the origin is invariant
under homotopy (i.e. under the smooth variation mentioned
above). Hence gauge invariance in the presence of boundaries
may be proved by combining this result with a hard WKB analysis
of coupled eigenvalue equations. Remarkably, one would then
obtain yet another application of the Atiyah-Patodi-Singer
theory of Riemannian 4-geometries with boundary [4].
\vskip 0.3cm
\noindent
(5.7) A mode-by-mode analysis of linearized gravity in the
presence of boundaries in the de Donder gauge has just been
completed, including gauge modes and ghost modes. Again, on
taking a flat Euclidean background bounded by two concentric
3-spheres, the mode-by-mode analysis of Faddeev-Popov quantum
amplitudes agrees with the result of covariant Schwinger-DeWitt
formalism [12].

The developments presented in sections 3 and 4 may lead to a
deeper understanding of conformal anomalies and gauge
invariance in quantum field theory. Hence we think that Henri
Poincar\'e would be pleased, if not surprised, to see how many
applications of asymptotic analysis are relevant for modern
quantum field theory.
\vskip 1cm
\leftline {\bf Acknowledgments}
\vskip 1cm
\noindent
I am indebted to the Italian MURST for financial support to
attend the Henri Poincar\'e Conference in Protvino.
\vskip 1cm
\leftline {\bf References}
\vskip 1cm
\item {[1]}
Poincar\'e H. (1886) {\it Acta Mathematica} {\bf 8}, 295.
\item {[2]}
Esposito G. (1994) {\it Quantum Gravity, Quantum Cosmology
and Lorentzian Geometries}, Lecture Notes in Physics, New
Series m: Monographs, Vol. m12, second corrected and enlarged
edition (Berlin: Springer-Verlag).
\item {[3]}
Seeley R. T. (1967) {\it Amer. Math. Soc. Proc. Symp. Pure Math.}
{\bf 10}, 288.
\item {[4]}
Atiyah M. F., Patodi V. K. and Singer I. M. (1976)
{\it Math. Proc. Camb. Phil. Soc.} {\bf 79}, 71.
\item {[5]}
Moss I. G. (1989) {\it Class. Quantum Grav.} {\bf 6}, 759.
\item {[6]}
D'Eath P. D. and Esposito G. (1991) {\it Phys. Rev.}
{\bf D 43}, 3234.
\item {[7]}
Barvinsky A. O., Kamenshchik A. Yu. and Karmazin I. P. (1992)
{\it Ann. Phys.} {\bf 219}, 201.
\item {[8]}
Kamenshchik A. Yu. and Mishakov I. V. (1993) {\it Phys. Rev.}
{\bf D 47}, 1380.
\item {[9]}
Kamenshchik A. Yu. and Mishakov I. V. (1994) {\it Phys. Rev.}
{\bf D 49}, 816.
\item {[10]}
Esposito G. (1994) {\it Class. Quantum Grav.} {\bf 11}, 905.
\item {[11]}
Esposito G., Kamenshchik A. Yu., Mishakov I. V. and
Pollifrone G. (1994)
{\it Class. Quantum Grav.} {\bf 11}, 2939.
\item {[12]}
Esposito G., Kamenshchik A. Yu., Mishakov I. V. and
Pollifrone G. (1994) {\it Phys. Rev.} {\bf D 50}, 6329.

\bye